\documentclass{emulateapj}
\usepackage{apjfonts}
\usepackage{psfig}
\usepackage{mathrsfs}
\usepackage{amsmath}

\slugcomment{}

\shorttitle{Radial extent of the SGB in NGC~1851}
\shortauthors{Zoccali et al.}

\begin{document}

\title{The radial extent of the double subgiant branch in NGC~1851
\footnote{Observations collected at the European Southern Observatory,
Paranal, Chile (ESO program 68.D-0510A).}}

\author{M. Zoccali\altaffilmark{1}}
\author{E. Pancino\altaffilmark{2}}
\author{M. Catelan\altaffilmark{1,3,4}}
\author{M. Hempel\altaffilmark{1}}
\author{M. Rejkuba\altaffilmark{5}}
\author{R. Carrera\altaffilmark{2}}

\altaffiltext{1}{P. Universidad Cat\'olica de Chile, Departamento de Astronom\'\i a y 
Astrof\'\i sica, Casilla 306, Santiago 22, Chile; email: mzoccali, mcatelan, 
mhempel@astro.puc.cl} 

\altaffiltext{2}{INAF - Osservatorio Astronomico di Bologna, Via Ranzani 1, I-40127 
Bologna, Italy; email: elena.pancino@oabo.inaf.it}

\altaffiltext{3}{John Simon Guggenheim Memorial Foundation Fellow}

\altaffiltext{4}{On sabbatical leave at Catholic University of America, 
Department of Physics, 200 Hannan Hall, Washington, DC 20064}

\altaffiltext{5}{European Southern Observatory, Karl-Schwarzschild-Strasse 2, D-85748 
Garching, Germany; email: mrejkuba@eso.org}

\begin{abstract}
Recent  HST-ACS observations  revealed the  presence of  a  double 
subgiant  branch (SGB)  in  the  core of  the  Galactic globular  cluster
NGC~1851. This  peculiarity was tentatively explained  by the presence
of a second  population with either an age difference  of about 1 Gyr,
or a  higher C+N+O abundance, probably  due to pollution  by the first
generation of stars.  

In the present Letter, we  analyze VLT-FORS $V,I$ images, covering 
$12.7\arcmin \times 12.7 \arcmin$, 
in the southwest quadrant of the cluster, allowing us
to  probe the  extent of  the double  SGB from  $\sim$1.4  to $\sim$13
arcmin  from the  cluster center.   Our study  reveals, for  the first
time,  that the  ``peculiar'' population  is the  one associated  to the
fainter SGB.  Indeed,  while the percentage of stars  in this sequence
is about 45$\%$ in the cluster  core (as previously found on the basis
of HST-ACS data), we find that  it drops sharply, to a level consistent
with zero  in our data, at  $\sim$2.4 arcmin from  the cluster center,
where  the brighter  SGB, in  our  sample, still  contains $\sim$  100
stars.  Implications for the proposed scenarios are discussed.

\end{abstract}


%___________________________________________________________________________
\section{Introduction}

Until 2004, $\omega$~Centauri (NGC~5139) and M54 (NGC~6715) used to be
the only two  globular clusters (GCs) with clear  evidence of multiple
stellar  components, based  on the  presence  of a  complex red  giant
branch      (RGB)      in      their     color-magnitude      diagrams
\citep[CMDs;][]{epea00,msea07}.  Thanks to the exquisite image quality
of  the  ACS  camera  on  board  HST, complex  CMD  features  such  as
splittings and  multimodalities were confirmed in  these clusters, and
even subtler features have been detected in other GCs.  \citet{lbea04}
discovered that the main sequence (MS) of $\omega$~Cen is also bimodal
\citep[see also][]{svea07},  and \citet{gpea05} demonstrated  that the
blue MS  was more metal-rich than  the red one, and  suggested that an
exceptionally  high  helium content  ($Y\sim0.38$)  could explain  its
bluer color.   A triple MS was discovered  in NGC~2808 \citep{gpea07},
while NGC~1851,  and possibly NGC~6388, have  double subgiant branches
\citep[SGB;][]{amea08a,gp08}.  In addition,  several LMC clusters show
evidence   of   multiple    sequences,   mostly   along   their   SGBs
\citep{amea08b}.

Interestingly, the  CMD sequences  in these GCs  tend to  be discrete,
pointing to multiple, but  separated, generations of stars (as opposed
to  extended  star  formation  periods)  or to  multimodality  in  the
chemical  composition (as opposed  to metallicity  spreads).  However,
each  of the  clusters mentioned  shows a  different  CMD peculiarity,
resulting from  a different enrichment/formation  history \citep[see][
for a review]{gp08}.

Some  of the  observed features,  such as  the double  SGBs,  could be
explained  in  terms of  age  differences  between the  subpopulations
\citep{amea08a}. However,  a star formation history  consisting of two
bursts with  an age gap  but identical composition would  be unlikely,
due to the  winds of massive stars and  core-collapse supernovae (SNs)
of the  first generation quickly enriching the  medium, hence altering
the  initial  composition  of  the  material  from  which  the  second
generation is  born.  Therefore, the most  plausible scenarios involve
multiple populations born at different times with different abundances
\citep{scea08,msea08,ar08}.   The presence  of two  coeval populations
with different chemistry might be  possible, in principle, in the {\it
pollution}  scenario, where  some  of the  stars  have their  surfaces
contaminated  by  the winds  of  stars  belonging  to the  same  burst
\citep[e.g.,][]{fdea83}.  However, as  discussed by \citet{ar08}, this
would  lead to  spreads rather  than multimodalities,  given  that the
amount  of  contamination would  be  a  continuous  function of  mass,
velocity, and orbit for each star inside the cluster.

In the  present paper we use  ground-based data to  analyze the radial
behavior of the double  SGB identified by \citet{amea08a} in NGC~1851.
This      is     a      massive      cluster     \citep[$3\times10^5\,
M_\odot$;][]{mclvdm05},    previously   studied   from    the   ground
\citep[e.g.,][]{aw92,isea98,mbea01}, whose main known peculiarity is a
bimodal  horizontal branch  (HB).  The  ACS study  by \citet{amea08a},
confined  to the cluster  core, showed  that the  SGB splits  into two
similarly populated sequences, with 45$\%$ of the stars in the fainter
SGB  and  55$\%$  in  the  brighter one.   \citet{scea08}  proposed  a
different C+N+O content as a  possible explanation for the double SGB,
while \citet{msea08}  and \citet{mc09} claimed that  such a difference
should not be  coupled with a variation in the  helium content $Y$, or
else the observed HB cannot be reproduced. By means of high-resolution
spectroscopy of 8 giants in NGC~1851, \citet{yg08} claimed the absence
of any spread  or bimodality in [Fe/H], together  with the presence of
large, correlated star-to-star variations  in the abundances of O, Na,
and Al.

The  ACS  data   cover  only  the  cluster  core,   while  the  quoted
ground-based photometries for NGC~1851  did not allow to constrain the
shape of the  SGB with the precision required to  detect an SGB split.
The  present paper  is the  first attempt  at constraining  the radial
extent of the two populations associated with the double SGB.

%_____________________________________________________________________
%-------
\begin{figure}[ht]
\includegraphics[angle=0,width=9.0 cm]{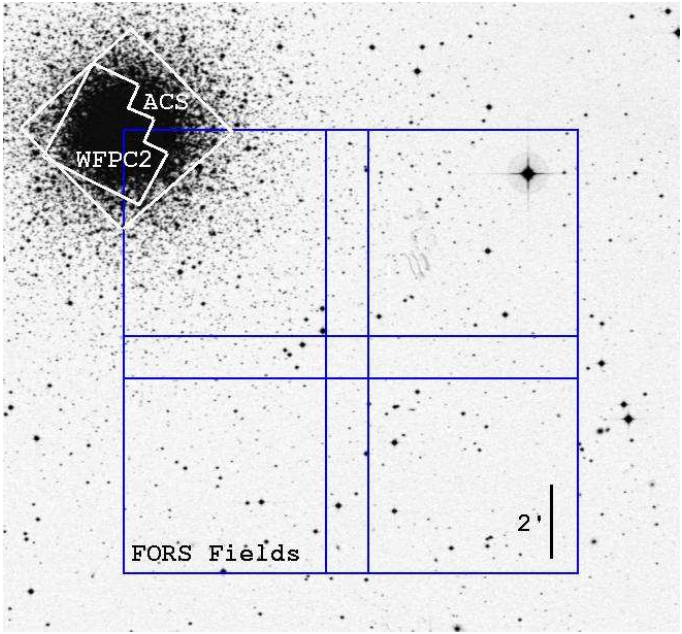}
\caption{DSS image of NGC~1851 with the position of the VLT-FORS 
$2\arcmin \times 2\arcmin$
mosaic. The HST-WFPC2 and HST-ACS fields 
are also 
indicated. North is up and 
east to the left.}
\label{fields}
\end{figure}
%-------

\section{Observations and Data Analysis}

Our data comes from the  pre-imaging run of a spectroscopic survey for
CN-CH  anticorrelations  in several  GCs.   The  NGC~1851 images  were
acquired on September 28$^{\rm th}$,  2001, in service mode, using the
FORS2-VLT  2k$\times$2k  SiTE  CCD.   The  4-field mosaic  in  the  SW
quadrant of the cluster  (Fig.~\ref{fields}) contains a short (2s) and
a long (15s) exposure taken with each of $V$ and $I$ at each position.
There is $\sim$15\% overlap  between adjacent fields, yielding a total
imaged area $12.7\arcmin \times 12.7 \arcmin$.  The average seeing was
$1.3 \arcsec$.

The images were  de-biased and flat-fielded by the  ESO FORS pipeline.
PSF-fitting photometry  was carried  out with DAOPHOTII,  ALLSTAR, and
ALLFRAME \citep{pbs87,pbs94}, using a constant PSF across the field~--
which empirically yielded the best results.  Though no standard fields
were  observed,  the  photometry   was  calibrated  by  means  of  the
$\sim$1200 stars  in common with  the \citet{mbea01} catalogue,  for a
$8\arcmin \times  8\arcmin$ field centered  on the cluster.   Only the
innermost  FORS field  had stars  in common  with this  catalogue; the
other  three fields  were calibrated  thanks to  the  relatively large
overlap    with   the    central   one,    and    amongst   themselves
(Fig.~\ref{fields}).   While  crucial to  calibrate  our  data to  the
standard Johnson-Cousins system, the photometry by \citeauthor{mbea01}
does  not reach  the  required  precision to  assess  the presence  or
absence of a double SGB.

% Fig.2 and relative text removed from here.

In order to improve the definition of the SGB, a selection was applied
based  on  the  photometric errors,  as  well  as  the CHI  and  SHARP
parameters,  obtained in  the PSF  fitting process  \citep{pbs87}. The
smallest  photometric  error  for  each  star  is  the  Poisson  error
associated  to   its  flux.   Indeed,   the  lower  envelope   of  the
distribution  of photometric  errors yielded  by DAOPHOT  is  always a
smooth  function of  the magnitude.   In  order to  select stars  with
unusually large errors {\it for  their magnitude}, we fitted the lower
envelope of the error distribution, and rejected stars having an error
in excess of 0.03~mag in $V$ or $I$ with respect to this envelope.  In
addition,  selection  limits  of  CHI$<2.5$  and  $-1<$SHARP$<1$  were
imposed on  all stars.  The procedure  is similar to the  one used in,
e.g.,  \citet[][  their  Fig.~5]{mzea09}.   We  emphasize  that  these
selection criteria  are very relaxed:  only a small percentage  of the
stars in the turnoff-SGB region were rejected.

%-------
\begin{figure}[ht]
\includegraphics[angle=0,width=9.2 cm]{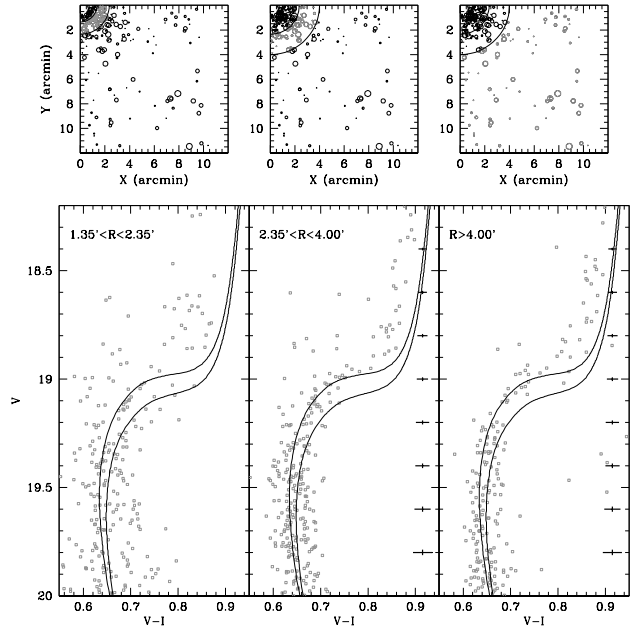}
\caption{Bottom: three radial selections  on the FORS CMD (see labels,
with numbers in  arcmin) corresponding to the stars  marked in gray
in the maps above each panel. The  solid lines are two isochrones for 
10 and 11~Gyr, with normal $\alpha$-enhanced composition, and metallicity 
$Z=0.002$. Formal photometric errors, as given by DAOPHOT are shown on 
the right side (middle and right panels).}
\label{cmdiso}
\end{figure}
%-------

\section{The spatial distribution of SGB stars}

Figure~\ref{cmdiso}  shows  the   error-selected  and  mosaic-ed  FORS
photometry, in  three radial annuli.  The photometry  in the innermost
annulus, at  a distance  between $1.35\arcmin$ and  $2.35\arcmin$ from
the center,  is rather poor, due  to the high stellar  density and the
rather  poor  seeing.  Also  shown  are  the  two isochrones  used  by
\citet{scea08}  to  fit  the  HST-ACS SGB,  having  Z=0.002,  standard
$\alpha$-element enhancement  and helium content  ($Y=0.25$), and ages
of 10  and 11 Gyr.   The adopted distance  modulus [(m$-$M)$_0=15.40$]
and reddening [E($B-V$)=0.06] are  constrained by the middle and right
panels, assuming (rather arbitrarily at  this point) that the SGB seen
in those CMDs corresponds to the brighter one (see below).

Although  the SGB  in  the  innermost region  is  poorly defined,  its
appearance  is   compatible  with  the  presence   of  two  sequences,
approximately  matched by  the isochrones.   All the  stars  above the
younger isochrone with colors $V\!-\!I \lesssim 0.8$ are likely blends
(see below).  The middle and right panels show the CMD relative to the
region between  $2.35\arcmin$ and $4\arcmin$,  and outside $4\arcmin$,
respectively.  Only one of the two SGBs is well populated in both.  If
the  SGB were double,  this would  be readily  apparent in  this plot,
given that the spread in magnitude, likely due entirely to photometric
errors,  is  smaller than  the  expected  separation  between the  two
sequences.\footnote{Note that the lower RGB is not well matched by the
isochrones.    This  is   a  known   problem,  due   to   an  improper
transformation  to the  observational plane  in these  bands (Cassisi,
private communication).   Indeed, the isochrone fit  shown in Figure~2
of \citet{scea08} does not include the lower RGB.}

%-------
\begin{figure}[ht]
\includegraphics[angle=-90,width=9.2 cm]{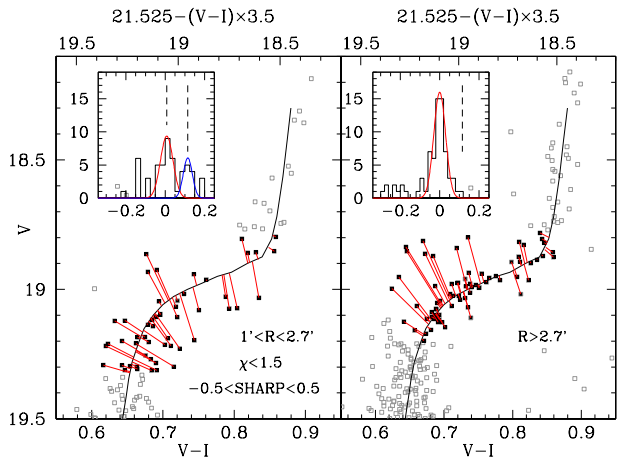}
\caption{Projection of  the SGB stars on  the mean ridgeline, in the
inner   (left)   and  outer   (right)   cluster   regions.  The   color
transformation shown  in the  top label has  been applied so  that the
color and  magnitude baselines are  equal, and a  meaningful projection
onto  the ridgeline  can be  carried out.   The distributions  of the
distances of  SGB stars from  the ridgeline  are shown in  the figure
insets. Vertical dashed  lines mark the expected positions  of the two
SGBs.}
\label{gauss}
\end{figure}
%-------

In  order  to  better  quantify  the  visual  impression  provided  by
Figure~\ref{cmdiso}, we  analyze the distribution of  stars around the
mean ridgeline  in Figure~\ref{gauss}.  Following  a procedure similar
to that  in \citet{gpea99}, a coordinate transformation  is applied to
the x-axis, in  order to give equal weight to  color and magnitude and
thus  correctly   project  each  star  on  the   fiducial  line.   The
distribution of the distances of SGB stars from the ridgeline is shown
in the figure  insets. A strong selection on CHI  and SHARP applied to
the photometry  of the inner  region confirmed our suspicion  that the
bright blue stars above the turnoff are blends.\footnote{The different
CHI,SHARP selection  criteria in the two  panels of Figure~\ref{gauss}
are justified  by the argument that  in the inner region  a double SGB
remains even  after a  strong selection,  hence it is  not due  to the
presence of  blends.  On the  contrary, no double  SGB is seen  in the
outer part,  even with  a very loose  selection.}  In the  inner region
there are  indications of the presence  of two SGBs,  separated by 0.1
mag, as  seen with  ACS.  Two Gaussians  fit this  distribution better
than a single one, with 88.2$\%$ confidence, according to the KMM test
\citep{abz94}. However,  the same test  indicates, at better  than the
99.99$\%$ confidence  level, that a single Gaussian  provides a better
fit  to the SGB  distribution further  out.  The  sigma of  the latter
Gaussian  (0.03 mag)  is a  realistic estimate  of the  errors  of our
photometry, and is significantly  smaller than the expected separation
between the two SGBs (0.1 mag).

Clearly, the present photometry in  the inner region is not comparable
with the  ACS one.  However,  if a double  SGB were present in  the SW
quadrant  beyond $2.7\arcmin$, we  would certainly  have been  able to
detect  it.  In  what follows,  we will  carry out  two more  tests to
further examine this conclusion.

\subsection{Could the double SGB show up only in the F606W filter?}

Our CMD was obtained using $V,I$, unlike \citeauthor{amea08a}'s, which
used F606W,~F814W.  Since F606W  is significantly broader than Johnson
$V$, it  could in principle  include some spectral feature  (e.g., the
red CN  band) not included in  $V$. If this  feature differed markedly
between the  two SGBs, one might  see the SGB splitting  only in F606W
(ACS), and not in $V$ (FORS).

\begin{figure}[ht]
\includegraphics[angle=0,width=9cm]{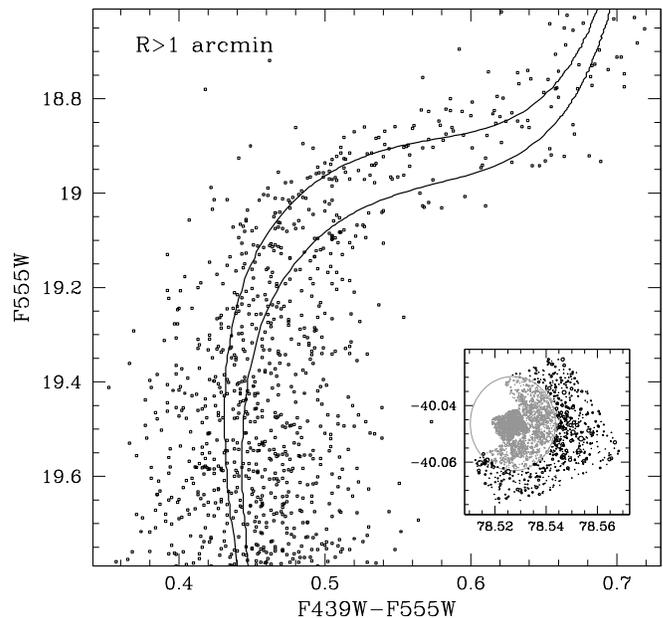}
\caption{Bottom: zoom  around the  SGB  region in  the HST-WFPC2  CMD.
  Solid  lines are  the same  isochrones as in  Fig.~\ref{cmdiso}, but
  transformed to the HST filter system. The radial interval considered 
  is shown in black in the inset map. }
\label{wfpc2}
\end{figure}

This  possibility can be  rejected based  on the  following arguments.
First,  if  the   SGB  splitting  were  due  to   either  age  or  CNO
overabundance,  it would  be  present also  in  luminosity, as  easily
verified  in the  isochrones computed  by \citet{scea08}.   This means
that the splitting would be  due to a different evolutionary path, and
not to different  line blanketing.  While the latter  could be present
preferentially in  the filter including  CN bands, the  former affects
all filters in a similar way.  Second, while the CMD in the left panel
of  Figure~\ref{gauss}  has too  few  stars  to  firmly establish  the
presence  of  a  double  SGB,  the CMD  from  the  HST-WFPC2  Snapshot
Project\footnote{The     data    are    publicly     available    from
http://dipastro.pd.astro.puc.cl/globulars/databases/snapshot/snapshot.html.}
\citep{gpea02} does show  it rather unambiguously.  Figure~\ref{wfpc2}
shows a zoom of the CMD  of NGC~1851, as obtained with HST-WFPC2, in a
region outside $1\arcmin$ from  the cluster center.  This demonstrates
that, whatever  the cause of the  SGB splitting, it is  visible in the
F555W filter, which is very similar to Johnson $V$.

%-------
%-------

\subsection{Which of the two SGBs disappears in the outer region?}

The  CMD in  Figure~\ref{wfpc2} also  shows  that the  fainter SGB  is
significantly  less  populated  than  the  brighter  one,  the  former
including  $\sim 40\%$  as many  stars as  the latter.   This strongly
suggests that  the ``peculiar'' population, intended here  as the less
numerous one, corresponds to the fainter SGB.

%-------
\begin{figure}[ht]
\includegraphics[angle=-90,width=9.5cm]{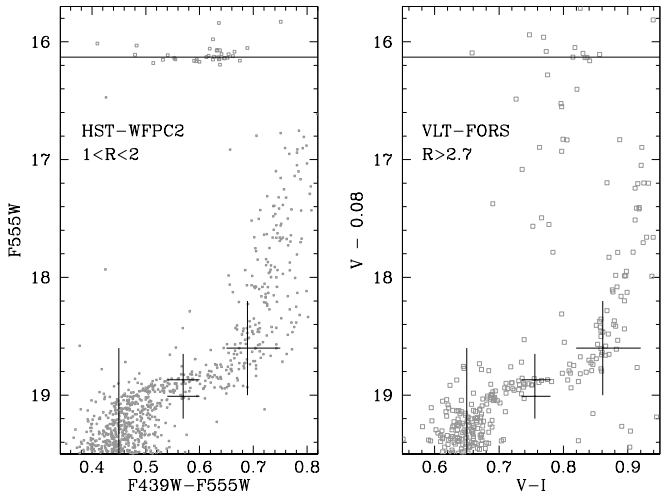}
\caption{WFPC2  (left) and  FORS (right) CMDs. The
FORS magnitudes  have been  shifted upwards by  0.08 mag, in  order to
match the level of the WFPC2 red HB. The vertical lines mark the color
of the MS  at $V=F555W=19$, the RGB at  $V=F555W=18.6$, and their mean
value in each  CMD. The two horizontal marks in the  middle of the SGB
mark the magnitude of the brighter and fainter SGBs in both diagrams.}
\label{sgb_p}
\end{figure}
%-------

To further check this, we shifted the FORS CMD vertically
by   0.08  mag,   so  that   its  red   HB  matches   the   WFPC2  one
(Fig.~\ref{sgb_p}).   This  removes the  small  zero point  difference
between the Johnson $V$ and  HST F555W filters, while the color term
is negligible  given the similarity between the  two filter passbands.
The single SGB visible in the  FORS CMD can be identified in the WFPC2
CMD by means of its magnitude  at a fixed (arbitrary) color. Since the
colors  on the  x-axis are  different in  the two  panels, we  need to
select a reference  position on the CMD, instead  of an absolute 
color value.  The color  in  the ``middle''  of the  SGB, i.e.,  the
average between the color of the MS at $V=F555W=19.5$ and the color of
the    RGB   at   $V=F555W=18.6$,    was   selected    as   reference.
Figure~\ref{sgb_p} clearly  shows that,  at this reference  color, the
brighter SGB  has $V=18.87$  in both CMDs,  while the fainter  one has
$V=19.01$ in  the WFPC2 data, and  is not visible in  the FORS CMD.

This  exercise clearly demonstrates  that the  ``peculiar'' population
corresponds to the  fainter SGB, which in our data  is present only in
the    central    $\sim   2\arcmin$~--    as    already   hinted    by
Figure~\ref{cmdiso}.  Further support to this idea comes from the fact
that \citet{amea08a} also  found the fainter SGB to  be more centrally
concentrated  than  the brighter  SGB  (see  their Fig.~6).   However,
having  data  only for  the  cluster  core,  they concluded  that  the
difference was  compatible with  the two groups  of stars  sharing the
same spatial  distribution. Instead, our study clearly  shows that the
fainter SGB is much more centrally concentrated.

%_____________________________________________________________________
\section{Discussion}

We have demonstrated that the ratio between faint and bright SGB stars
(fSGB/bSGB)  in NGC1851,  which is  45/55 in  the cluster  core, drops
dramatically in the outer region,  to a level consistent with zero, at
least  in the  cluster's SW  quadrant. We  shall now  use  the central
concentration of  the ``peculiar'' population associated  to the faint
SGB to pose constraints on its origin.

Three hypotheses  have been proposed  to explain the SGB  splitting in
NGC~1851, namely:  {\it i)}  A pure age  difference, by  about $\sim$1
Gyr; {\it ii)} A CNO overabundance  by about a factor of 2, associated
with the  brighter SGB~-- the latter  also being $\sim$  2 Gyr younger
than the  ``normal,'' fainter  SGB; and {\it  iii)} A  CNO enhancement
associated  with  the  fainter  SGB  at  a  fixed  age  (i.e.,  coeval
populations).

\citet{msea08}  suggested that the  progeny of  the fainter  SGB stars
should be found  both on the blue  and red sides of the  HB, while the
brighter SGB  should produce red HB  stars only.  By  coupling the SGB
shape with  the constraints  posed by star  counts along the  HB, they
excluded the brighter SGB population as the extreme, CNO-enhanced one.

{\bf  Pure age  difference}.   If  the fainter  SGB  corresponds to  a
population  $\sim$1  Gyr older  than  the  brighter  one, its  central
concentration would  imply something  akin to an  inside-out formation
scenario,  with a  first episode  of star  formation, confined  to the
center and converting into stars  only a small amount of the available
gas, followed by a second burst 1 Gyr later, converting into stars all
the  remaining gas,  and  involving the  whole  cluster volume.   This
galaxy-like  scenario seems  rather  unlikely for  a low-mass  stellar
system;  a  specific   dynamical  model  might  confirm/disprove  this
hypothesis on more quantitative grounds.

{\bf CNO enhancement of the  brighter SGB}.  If the brighter SGB stars
are CNO-enhanced,  then they  should be $\sim$2  Gyr younger  than the
normal-abundance,  fainter  SGB  ones.   However,  in  this  case  the
chemically normal population should  be the minority one, located only
in the center,  but it should somehow have  succeeded in enriching the
second-generation stars  throughout the cluster  volume.  Clearly this
is not a  plausible solution, because the relatively  small total mass
of  the primordial population  would not  be able  to enrich  the more
massive  second  generation.   This  scenario  was  also  excluded  by
\citet{msea08} on the basis of star counts along the HB.

{\bf CNO enhancement of the fainter SGB}. If the fainter SGB is coeval
with the brighter  SGB, then it should have  an extreme metal mixture,
characterized by CNO  enhancement, strong anticorrelations between C-N
and O-Na, but  the same $Z$ and $Y$  \citep{scea08}.  The present data
allow us to  identify such a scenario as the  most plausible one.  The
bulk  of the  stellar  population  in NGC~1851  would  have been  born
$\sim$12 Gyr  ago, and in the  timespan between 20-30 Myr  and 300 Myr
after its birth, the winds from its massive AGB stars would accumulate
enriched gas at the bottom of the potential well, i.e., at the cluster
center \citep{ar08}.  At some  point during this time, the accumulated
material could undergo a second episode of star formation, soon enough
that the age difference would not be detectable with the current data.

The present analysis allowed us to identify the fainter SGB population
as the  ``peculiar'' one, being much more  centrally concentrated than
the brighter SGB, and either  older or CNO enriched (and coeval within
the errors,  $\sim$0.4 Gyr).  Both  scenarios are compatible  with the
results    of     \citet{msea08}.     However,    in     the    middle
($2.35<R[\arcmin]<4.0$)  and outer ($R>4\arcmin$)  radial bins  of the
FORS data, the  stars on the blue and  red sides of the HB  are in the
proportion $\mathcal{B:R}=2:8$ and $2:3$, respectively. If the progeny
of the  brighter SGB can  evolve only to  the red HB, as  suggested by
\citeauthor{msea08}, then  the blue HB  should disappear in  the outer
region,  together  with the  fainter  SGB.   Despite the  small-number
statistics, we do  see 2 blue HB stars even out  to $R \sim 4\arcmin$,
where  the  fainter  SGB  has  disappeared  in  our  data.   Based  on
\citeauthor{aw92}'s  (\citeyear{aw92}) data,  covering  a wider  field
than the  SW quadrant studied  here, we count  $\mathcal{B:R}=8:18$ in
the  middle radial  bin.  If  confirmed  by statistically  larger
datasets, the  presence of some blue  HB stars over  the whole cluster
area  would be  in  contraddiction  with the  absence  of fainter  SGB
stars outside 2.4 arcmin.

% This confirms  that the blue  HB extends over
% the  whole  cluster  area,  contrary  to  the  fainter  SGB,  which~--
% according to our data~-- is not present outside $2.4\arcmin$ in the SW
% quadrant.

Before  closing,  we  note  that   it  is  not  obvious  how  multiple
populations  may remain  radially  segregated  in an  old  GC such  as
NGC~1851.  The central relaxation time of NGC~1851 is only $2.6 \times
10^7$~yr,  whereas at  the half-light  radius  it is  still just  $3.2
\times  10^8$~yr  \citep{sd93}.    Thus,  any  differences  in  radial
distribution  among  the  different  stellar  components  should  have
disappeared long  ago.  On the  other hand, we estimate  that, outside
the  half-light  radius  of  NGC~1851  ($\approx  1.8  \arcmin$),  the
relaxation  time becomes longer  than a  Hubble time  \citep[eq.  2-61
in][]{ls87}.  Since our observations cover mostly regions further out,
it is  conceivable that the  spatial distribution of  distinct stellar
populations  in   the  cluster   will  only  become   more  thoroughly
homogeneized in  the distant future.  Further studies  of the detailed
spatial  distribution  of  the  stars associated  with  the  different
sub-populations, along  with their chemical  composition, are strongly
encouraged.

\acknowledgments
This work has been partly funded by the FONDAP Center for Astrophysics 
15010003 and Proyecto Basal PFB-06/2007 (MZ and MC), by Proyecto 
FONDECYT Regular \#1085278 (MZ) and \#1071002 (MC), and by a 
John Simon Guggenheim Memorial Foundation Fellowship (MC). EP acknowledges
the support of ESO Vitacura, Chile, through the {\it ESO Science Visitor
Programme}.

%_____________________________________________________________________

\end{document}